\documentclass[preprint,showpacs,preprintnumbers,amsmath,amssymb]{article}

\usepackage{graphicx}
\usepackage{color}

\begin{document}

\title{Ba$\tilde{\rm n}$ados--Teitelboim--Zanelli Black Hole in the Information Geometry}
\author{Hiroaki Matsueda${}^{1}$\thanks{matsueda@sendai-nct.ac.jp} and Tatsuo Suzuki${}^{2}$\thanks{suzukita@sic.shibaura-it.ac.jp}
\\
\begin{small}
${}^{1}$Sendai National College of Technology, Sendai 989-3128, Japan 
\end{small}
\\
\begin{small}
${}^{2}$Department of Mathematical Sciences, 
Shibaura Institute of Technology, Saitama 337-8570, Japan
\end{small}
}
\date{}

\maketitle

\begin{abstract}
We examine the Ba$\tilde{\rm n}$ados--Teitelboim--Zanelli (BTZ) black hole in terms of the information geometry and consider what kind of quantum information produces the black hole metric in close connection with the anti-de Sitter space/conformal field theory (AdS/CFT) correspondence. We find a Hessian potential that exactly produces both the BTZ metric and the entanglement entropy formula for CFT${}_{1+1}$ at a finite temperature. Taking a free-falling frame near the event horizon is a key procedure to derive these exact results. We also find an alternative Hessian potential that produces the same BTZ metric, which is found using the duality relation based on the Legendre transformation. We realize that the dual representation originates from the entanglement Hamiltonian on the CFT side. Our results suggest that the present information-geometrical approach is very powerful for understanding the mechanism of the holographic renormalization group such as the AdS/CFT correspondence.
\end{abstract}


\section{Introduction}

The possible application of the information geometry to space-time physics has a long history. Since an information metric such as the Fisher metric can be defined from our target microscopic model, the resultant classical space-time that emerges from the information metric may answer fundamental scientific questions associated with quantum gravity, efficient quantum information storage, and so on. Unfortunately, most previous works seem to lack physical interpretation since they are based on purely mathematical or information-geometrical viewpoints regarding the amount of information, not physical motivations. However, one of the authors (HM) has found that some important aspects in the anti-de Sitter space/conformal field theory (AdS/CFT) correspondence in string theory can be well captured by the information geometry~\cite{Maldacena1,Maldacena2,Takayanagi1,Matsueda}. Therein, it is important to rely on the entanglement entropy scaling in the CFT. In this context, we would like to know more about the functionality of the information geometry by taking other well-known examples in the AdS/CFT.

Here we focus on the Ba$\tilde{\rm n}$ados--Teitelboim--Zanelli (BTZ) black hole~\cite{BTZ1,BTZ2}. The BTZ black hole geometry is the solution of the vacuum Einstein field equation in $(2+1)$ dimensions with a negative cosmological constant~\cite{BTZ1,BTZ2}. From the viewpoint of the AdS/CFT, its dual field theory is the $(1+1)$-dimensional CFT (CFT${}_{1+1}$), and the presence of the black hole corresponds to finite-temperature effects on the CFT. An interesting question is whether the information geometry can capture this holographic transformation. Our answer is yes, and we find the Hessian potentials to realize this transformation exactly.

Of particular importance among our present results is the existence of the duality relation. We find two different Hessian potentials that produce the same BTZ metric as well as the entanglement entropy formula for the dual quantum field theory. These potentials are related to each other by the Legendre transformation characteristic in the information geometry. The point behind this transformation is that it is equivalent to the duality in the AdS/CFT correspondence. Their surprising similarity is still a mysterious observation, but the present results are sufficient to realize the powerfulness of the information geometry in the examination of the AdS/CFT correspondence.

The organization of this paper is as follows. In the next section, we remark on basic notions of the information geometry. In Sect. 3, we explain the close relationship between the free-falling frame of the BTZ black hole and the entanglement entropy in CFT${}_{1+1}$. Our main results are shown in Sects. 4 and 5. We mention the duality relation naturally inherent in the information geometry and its close relationship to the AdS/CFT correspondence. Finally, we summarize our study.

\section{Basic Notions of the Information Geometry}

We first summarize basic notions of the information geometry that are frequently used throughout this paper~\cite{Amari,Shima}. Let us consider the probability distribution $\{\lambda_{n}\}$ with $\sum_{n}\lambda_{n}=1$. The distribution originates from the Schmidt coefficients for the quantum pure state
\begin{eqnarray}
\left|\psi\right>=\sum_{n}\sqrt{\lambda_{n}}\left|n\right>_{A}\otimes\left|n\right>_{\bar{A}}, \label{state}
\end{eqnarray}
where $\left|n\right>_{A}$ and $\left|n\right>_{\bar{A}}$ are basis states for a subsystem $A$ and its complement $\bar{A}$, respectively. They share the same index $n$ since these systems are entangled with each other. The entanglement spectrum $\gamma_{n}$ is defined by
\begin{eqnarray}
\lambda_{n}=e^{-\gamma_{n}}. \label{lambda}
\end{eqnarray}
We assume that the distribution has an exponential form defined by
\begin{eqnarray}
\gamma_{n}(\theta)=\psi(\theta)-\theta^{\alpha}F_{n,\alpha}. \label{gamma}
\end{eqnarray}
Here, $\theta^{\alpha}$ are called the canonical parameters, and they are determined from the parameters of the original quantum model that defines Eq.~(\ref{state}). In other words, we must first represent $\gamma_{n}$ by the form of Eq.~(\ref{gamma}) and find how the canonical parameters are related to the model parameters. For instance, one of the canonical parameters is obtained as $1/l^{2}$, where $l$ is the size of subsystem $A$ in the spatially one-dimensional lattice free-fermion model~\cite{Matsueda}. The function $\psi$ is called the Hessian potential, which plays crucial roles in the examination of the quantum/classical correspondence in terms of the information geometry. The assumption of the canonical form is natural since the partial trace causes finite-temperature-like effects on the spectrum. More precisely, by combining Eq.~(\ref{lambda}) with Eq.~(\ref{gamma}) as well as the normalization condition for $\lambda_{n}$, we obtain the following thermodynamic representations:
\begin{eqnarray}
\psi=\log Z \; , \; Z=\sum_{n}e^{\theta^{\alpha}F_{n,\alpha}} \; , \; \lambda_{n}=\frac{1}{Z}e^{\theta^{\alpha}F_{n,\alpha}}.
\end{eqnarray}
The average of $\gamma_{n}$ is simply the entanglement entropy
\begin{eqnarray}
S=\left<\gamma\right>=\sum_{n}\lambda_{n}\gamma_{n}.
\end{eqnarray}
When we define $\partial_{\alpha}=\partial/\partial\theta^{\alpha}$, we have
\begin{eqnarray}
\partial_{\alpha}\left<1\right>=-\sum_{n}\lambda_{n}\partial_{\alpha}\gamma_{n}=-\partial_{\alpha}\psi+\left<F_{\alpha}\right>=0.
\end{eqnarray}
Then, the entropy is represented as
\begin{eqnarray}
S=\psi-\theta^{\alpha}\left<F_{\alpha}\right>=\psi-\theta^{\alpha}\partial_{\alpha}\psi.
\end{eqnarray}

The geometric representation of Eq.~(\ref{state}) in terms of the information geometry is also possible by starting with Eq.~(\ref{gamma}). The Fisher metric, the leading order of the relative entanglement entropy, is defined by
\begin{eqnarray}
g_{\mu\nu}=\left<(\partial_{\mu}\gamma)(\partial_{\nu}\gamma)\right>, \label{g1}
\end{eqnarray}
and this is also equal to
\begin{eqnarray}
g_{\mu\nu}=\left<\partial_{\mu}\partial_{\nu}\gamma\right>. \label{g2}
\end{eqnarray}
The line element is calculated as
\begin{eqnarray}
ds^{2}=\kappa g_{\mu\nu}d\theta^{\mu}d\theta^{\nu},
\end{eqnarray}
where the overall contant $\kappa$ will be determined later so that we can introduce appropriate physical units. Readers may wonder why the number of $\gamma_{n}$ changes between Eqs.~(\ref{g1}) and (\ref{g2}): this is because the probability distribution depends on the canonical parameters. Actually, the equality between Eqs.~(\ref{g1}) and (\ref{g2}) is derived from $\partial_{\mu}\partial_{\nu}\left<1\right>=0$:
\begin{eqnarray}
0 &=& \partial_{\mu}\sum_{n}\partial_{\nu}\lambda_{n} \nonumber \\
&=& -\partial_{\mu}\sum_{n}\lambda_{n}\partial_{\nu}\gamma_{n} \nonumber \\
&=& \sum_{n}\lambda_{n}(\partial_{\mu}\gamma_{n})(\partial_{\nu}\gamma_{n})-\sum_{n}\lambda_{n}\partial_{\mu}\partial_{\nu}\gamma_{n} \nonumber \\
&=& \left<(\partial_{\mu}\gamma)(\partial_{\nu}\gamma)\right>-\left<\partial_{\mu}\partial_{\nu}\gamma\right>.
\end{eqnarray}
Equation (\ref{g2}) does not seem to show a general covariance, but the original definition in Eq.~(\ref{g1}) is clearly covariant. Thus, their discrepancy is fictitious and the covariance should be conserved. Actually, for the new coordinates, $x^{a}$, we have
\begin{eqnarray}
\left<\partial_{\mu}\partial_{\nu}\gamma\right>=\left<\frac{\partial^{2}\gamma}{\partial x^{a}\partial x^{b}}\right>\frac{\partial x^{a}}{\partial\theta^{\mu}}\frac{\partial x^{b}}{\partial\theta^{\nu}}+\left<\frac{\partial\gamma}{\partial x^{b}}\right>\frac{\partial^{2}x^{b}}{\partial\theta^{\mu}\partial\theta^{\nu}}, \label{g4}
\end{eqnarray}
and the second term vanishes since we find that
\begin{eqnarray}
\left<\frac{\partial\gamma}{\partial x^{b}}\right>=\sum_{n}\lambda_{n}\frac{\partial(-\log\lambda_{n})}{\partial x^{b}}=-\frac{\partial}{\partial x^{b}}\sum_{n}\lambda_{n}=0.
\end{eqnarray}
When we make a special frame $\theta^{\alpha}$ and assume Eq.~(\ref{gamma}), the Fisher metric is reduced to the Hessian form
\begin{eqnarray}
g_{\mu\nu}=\partial_{\mu}\partial_{\nu}\psi. \label{g3}
\end{eqnarray}
Hereafter we particularly focus on this representation.

Note that the Hessian potential form has some ambiguity since the metric is invariant for the transformation $\psi\rightarrow\psi+\delta\psi$ with the condition $\partial_{\mu}\partial_{\nu}(\delta\psi)=0$. When we expand $\delta\psi$ as a power series of the canonical parameters up to the second order, we have
\begin{eqnarray}
\delta\psi=A+B_{\alpha}\theta^{\alpha}+C_{\alpha\beta}\theta^{\alpha}\theta^{\beta}.
\end{eqnarray}
The change in the metric is given by
\begin{eqnarray}
\partial_{\mu}\partial_{\nu}(\delta\psi)=C_{\mu\nu}+C_{\nu\mu}.
\end{eqnarray}
This means that the tensor $C_{\mu\nu}$ is antisymmetric if this term vanishes. In this case, the additional term of the entropy, $\delta S=\delta\psi-\theta^{\alpha}\partial_{\alpha}(\delta\psi)$, is given by
\begin{eqnarray}
\delta S=A+\frac{1}{2}\left(C_{\mu\nu}+C_{\nu\mu}\right)\theta^{\mu}\theta^{\nu}=A.
\end{eqnarray}
Since the potential originates from the probability distribution, the constant $A$ is uniquely determined by the normalization condition for the distribution.

Our classical metric $g_{\mu\nu}$ is defined by the Hessian potential, which originates from the entanglement spectrum of our quantum state. Therefore, we can naturally examine the classical geometrical meaning of our quantum system. Our problem is to find the potential form $\psi$ and the canonical parameters $\theta^{\alpha}$ that produce $g_{\mu\nu}$ for the BTZ black hole.

In the information geometry, the duality is an important concept and it corresponds to the Legendre transformation. The dual parameters $\eta_{\alpha}$ to the original canonical ones are defined by
\begin{eqnarray}
\eta_{\alpha}=-\partial_{\alpha}\psi, \label{eta}
\end{eqnarray}
and the dual potential is defined by
\begin{eqnarray}
\varphi=\theta^{\alpha}(\partial_{\alpha}\psi)-\psi=-S. \label{varphi}
\end{eqnarray}
Then, the metric is given by
\begin{eqnarray}
g^{\alpha\beta}=\partial^{\alpha}\partial^{\beta}\varphi,
\end{eqnarray}
with $\partial^{\alpha}=\partial/\partial\eta_{\alpha}$, and the line element is calculated as
\begin{eqnarray}
ds^{2}=\kappa g^{\alpha\beta}d\eta_{\alpha}d\eta_{\beta}.
\end{eqnarray}

\section{Free-Falling Frame of the BTZ Black Hole}

The BTZ black hole is represented by the metric
\begin{eqnarray}
ds^{2}=\frac{R^{2}}{z^{2}}\left(-f(z)dt^{2}+\frac{dz^{2}}{f(z)}+dx^{2}\right),
\end{eqnarray}
where the factor $f(z)$ is defined by
\begin{eqnarray}
f(z)=1-\left(\frac{z}{z_{0}}\right)^{2}=1-az^{2}
\end{eqnarray}
and the position $z_{0}$ denotes the event horizon. The length scale $R$ is called the curvature radius. We consider the static case in which the angular momentum is zero. For the moment, we omit $R$ (and also $\kappa$) and take
\begin{eqnarray}
ds^{2}=\frac{1}{z^{2}}\left(-f(z)dt^{2}+\frac{dz^{2}}{f(z)}+dx^{2}\right). \label{BTZ}
\end{eqnarray}
Later we will consider the physical units by recovering both $R$ and $\kappa$.

In the information geometry, the metric is related to the second derivative of the entanglement entropy, and it is necessary to take appropriate coordinates so that the metric does not diverge at the event horizon. We introduce Eddington-Finkelstein-like (or Kruskal-Szekeres-like) coordinates for an observer falling into the black hole. For this purpose, let us find the null coordinate. We start with
\begin{eqnarray}
ds^{2}=\frac{1}{z^{2}}\left\{f(x)\left(-dt^{2}+\left(\frac{dz}{f(z)}\right)^{2}\right)+dx^{2}\right\}
\end{eqnarray}
and take
\begin{eqnarray}
dl=\frac{dz}{f(z)}. \label{l}
\end{eqnarray}
By integrating Eq.~(\ref{l}), we obtain
\begin{eqnarray}
l = \int\frac{dz}{1-(z/z_{0})^{2}} = z_{0}\tanh^{-1}\left(\frac{z}{z_{0}}\right)+C,
\end{eqnarray}
where $C$ is a constant. By this equation, we have defined a family of possible coordinates that satisfy the invariance of the line element. The constant $C$ can be set for convenience, and we set $C=0$. By using this free-falling frame $l$, we find that
\begin{eqnarray}
ds^{2}=\frac{1}{z_{0}^{2}}{\rm cosech}^{2}\left(\frac{l}{z_{0}}\right)\left(-dt^{2}+dl^{2}\right) + \frac{dx^{2}}{z^{2}},
\end{eqnarray}
and we can see that the fictitious coordinate singularity has been removed.

Here, an interesting observation is
\begin{eqnarray}
\frac{1}{z_{0}^{2}}{\rm cosech}^{2}\left(\frac{l}{z_{0}}\right)
= -\partial_{l}\partial_{l}\log\left(\frac{z_{0}}{\epsilon}\sinh\left(\frac{l}{z_{0}}\right)\right).
\end{eqnarray}
This agrees well with the entanglement entropy for CFT${}_{1+1}$ at a finite temperature~\cite{Holzhey,Calabrese1,Calabrese2,Calabrese3,Hubeny}. Actually, the scaling formula of the entropy is given by
\begin{eqnarray}
S_{EE}=\frac{c}{3}\log\left(\frac{\beta}{\pi\epsilon}\sinh\left(\frac{2\pi l}{\beta}\right)\right), \label{formula}
\end{eqnarray}
where $c$ is the central charge and $\epsilon$ is the UV cutoff. For $\beta\rightarrow\infty$, we obtain the well-known zero-temperature formula $S=(c/3)\log(L/\epsilon)$. We can identify $L=2l$ with the subsystem size and also identify the event-horizon position $z_{0}$ with the inverse temperature
\begin{eqnarray}
z_{0}=\frac{\beta}{2\pi}. 
\end{eqnarray}
The relation between $z_{0}$ and $\beta$ is also consistent with field-theoretical analysis such as the finite-temperature multiscale entanglement renormalization ansatz~\cite{BTZ1,Matsueda2}. The second derivative indicates that the entanglement entropy is closely related to the Hessian potential except for the minus sign. Since $l$ itself is not a canonical parameter, more sophisticated treatment is necessary. However, this observation appears to be crucial since the general coordinate transformation to the regular coordinates is a key procedure holographically connecting the black hole thermodynamics with the entanglement entropy (see also Appendix A). This point will be precisely discussed in the next section.

\section{Exact Canonical Parameters and Hessian Potential for the BTZ Metric}

We examine the exact Hessian potentials for the BTZ metric in this and the next sections. Physically, we start with an important fact that the logarithmic entropy scaling is a generic feature of the CFT, and we consider that the second derivative of the logarithmic function can produce the warp factor of the AdS space. We thus obtain the BTZ result by generalizing these useful observations with the help of entanglement thermodynamics. In this paper, we mainly discuss physical viewpoints, and a more mathematically rigorous treatment and the dual structure of the information geometry will be presented elsewhere.

Before going into details, it is helpful to determine the canonical parameters of the well-known Gaussian distribution, which produces a hyperbolic geometry similar to the AdS metric. The probability distribution is given by
\begin{eqnarray}
p(X)=\frac{1}{\sqrt{2\pi}\sigma}\exp\left\{ -\frac{(X-\mu)^{2}}{2\sigma^{2}} \right\}, \label{p}
\end{eqnarray}
where $\mu$ is the mean value and $\sigma^{2}$ is its variance. We transform $p(X)$ into
\begin{eqnarray}
p(X)=\exp\left\{ \theta^{\alpha}F_{\alpha}(X)-\psi(\theta) \right\}
\end{eqnarray}
and compare it with Eq.~(\ref{p}). By assuming
\begin{eqnarray}
F_{1} &=& X , \\
F_{2} &=& -\frac{1}{2}X^{2} , 
\end{eqnarray}
we obtain
\begin{eqnarray}
\psi &=& \log\left(\sqrt{2\pi}\sigma\right)+\frac{1}{2}\left(\frac{\mu}{\sigma}\right)^{2}, \label{p00} \\
\theta^{1} &=& \frac{\mu}{\sigma^{2}} , \\
\theta^{2} &=& \frac{1}{\sigma^{2}} .
\end{eqnarray}
We invert $\theta^{1}$ and $\theta^{2}$ to obtain $\mu$ and $\sigma$:
\begin{eqnarray}
\mu &=& \frac{\theta^{1}}{\theta^{2}} , \\
\sigma &=& \frac{1}{\sqrt{\theta^{2}}} ,
\end{eqnarray}
and then
\begin{eqnarray}
\psi = -\frac{1}{2}\log\theta^{2}+\frac{1}{2}\log 2\pi+\frac{1}{2}\frac{(\theta^{1})^{2}}{\theta^{2}}.
\end{eqnarray}
By using these parameters, we can derive the hyperbolic metric
\begin{eqnarray}
ds^{2} &=& \frac{1}{\theta^{2}}(d\theta^{1})^{2}-2\frac{\theta^{1}}{(\theta^{2})^{2}}d\theta^{1}d\theta^{2} \nonumber \\
&& + \left\{ \frac{1}{2}\frac{1}{(\theta^{2})^{2}}+\frac{(\theta^{1})^{2}}{(\theta^{2})^{3}}\right\}(d\theta^{2})^{2} \nonumber \\
&=& \frac{d\mu^{2}+2d\sigma^{2}}{\sigma^{2}}.
\end{eqnarray}
Note that the logarithmic term of the potential is crucial to derive the warp factor $\sigma^{-2}$. The information entropy derived from Eq.~(\ref{p00}) is given by
\begin{eqnarray}
S &=& -\frac{1}{2}\log\theta^{2}+\frac{1}{2}\log(2\pi) + \frac{1}{2} \nonumber \\
&=& \log\sigma + \frac{1}{2}\log(2\pi) + \frac{1}{2}. \label{S00}
\end{eqnarray}
Note that this formula seems to be closely related to the logarithmic formula for the entanglement entropy. Let us consider the harmonic oscillator
\begin{eqnarray}
H=-\sigma\frac{d^{2}}{dX^{2}}+\frac{(X-\mu)^{2}}{4\sigma^{3}}.
\end{eqnarray}
By solving the Schr\"{o}dinger equation $H\Psi(X)=E\Psi(X)$, the ground state, $\psi_{0}(X)$, is given by
\begin{eqnarray}
\psi_{0}(X)=\sqrt{\frac{1}{\sqrt{2\pi}\sigma}}e^{-(X-\mu)^{2}/4\sigma^{2}},
\end{eqnarray}
and then the probability distribution in Eq.~(\ref{p}) is obtained as $p(X)=|\psi_{0}(X)|^{2}$. Here, the parameter $\sigma$ corresponds to the spatial region $l$ in which our quantum particle is roughly confined. In this sense, it is obvious that Eq.~(\ref{S00}) originates from the logarithmic formula of the entanglement entropy, even if we do not explicitly consider any spatial truncation for the environmental degrees of freedom. Note that our length scale $\sigma$ is measured in the flat space-time.

The BTZ metric can be derived by modification of the above Gaussian formula. Let us take the space-time and radial coordinates as
\begin{eqnarray}
t &=& \left(\frac{\theta^{0}}{\theta^{2}-a}\right)^{p}, \label{time} \\
x &=& \frac{\theta^{1}}{\theta^{2}}, \\
z &=& \frac{1}{\sqrt{\theta^{2}}}, \label{z}
\end{eqnarray}
with a nonzero constant $p$. The parameters $x$ and $z$ are direct extensions of $\mu$ and $\sigma$, respectively. Thus, the classical space coordinate is mapped onto the center position of a wave packet on the quantum side, and $z$ corresponds to an energy scale characterized by the spatial extension of the wave packet. We show in Appendix B how to determine the time coordinate by solving the Codazzi equations $\partial_{\lambda}g_{\mu\nu}=\partial_{\mu}g_{\lambda\nu}$. Since $x$ and $z$ are assumed, we cannot exclude other possibilities of $x$ and $z$ that also satisfy the Codazzi equations. To keep the space-time symmetry in the case of $z_{0}\rightarrow\infty$ ($a\rightarrow 0$), we assume $p=1$, and then
\begin{eqnarray}
t=\frac{\theta^{0}}{\theta^{2}-a}=\frac{z^{2}}{1-az^{2}}\theta^{0},
\end{eqnarray}
where $t\ge 0$ for $\theta^{0}\ge 0$ and $z<z_{0}$. An interesting observation is that $t\rightarrow\infty$ for $z\rightarrow z_{0}$. In other words, our time evolution becomes very slow near the event horizon.

Using the canonical parameters, the BTZ metric is transformed into
\begin{eqnarray}
ds^{2} &=& \frac{1}{z^{2}}\left( -f(z)dt^{2}+\frac{dz^{2}}{f(z)}+dx^{2} \right) \nonumber \\
&=& -\frac{1}{\theta^{2}-a}(d\theta^{0})^{2} + 2\frac{\theta^{0}}{(\theta^{2}-a)^{2}}d\theta^{0}d\theta^{2} \nonumber \\
&& + \frac{1}{\theta^{2}}(d\theta^{1})^{2} -2\frac{\theta^{1}}{(\theta^{2})^{2}}d\theta^{1}d\theta^{2} \nonumber \\
&& + \left\{\frac{1}{4}\frac{1}{\theta^{2}(\theta^{2}-a)}+\frac{(\theta^{1})^{2}}{(\theta^{2})^{3}}-\frac{(\theta^{0})^{2}}{(\theta^{2}-a)^{3}}\right\}(d\theta^{2})^{2}, \nonumber \\
\label{iemetric}
\end{eqnarray}
which can be derived from the following Hessian potential:
\begin{eqnarray}
\psi&=&\frac{1}{4a}\left\{ (\theta^{2}-a)\log(\theta^{2}-a)-\theta^{2}\log\theta^{2}\right\} \nonumber \\
&& +\frac{1}{2}\frac{(\theta^{1})^{2}}{\theta^{2}}-\frac{1}{2}\frac{(\theta^{0})^{2}}{\theta^{2}-a} \nonumber \\
&=& \frac{1}{4}\log\left(\frac{z^{2}}{1-az^{2}}\right)+\frac{1}{4}\frac{1}{az^{2}}\log\left(1-az^{2}\right) \nonumber \\
&& + \frac{1}{2}at^{2} + \frac{1}{2}\frac{x^{2}-t^{2}}{z^{2}}. \label{new}
\end{eqnarray}
The information entropy is then obtained as
\begin{eqnarray}
S(\theta)&=&\psi(\theta)-\theta^{\alpha}\partial_{\alpha}\psi(\theta) \nonumber \\
&=& -\frac{1}{4}\log(\theta^{2}-a)-\frac{1}{2}a\left(\frac{\theta^{0}}{\theta^{2}-a}\right)^{2} \nonumber \\
&=& \frac{1}{4}\log\left(\frac{z^{2}}{1-az^{2}}\right) -\frac{1}{2}at^{2}. \label{ie}
\end{eqnarray}
Equations (\ref{iemetric}) and (\ref{ie}) provide us with a holography relation between the classical geometry and a quantum system characterized by the entropy. They are correlated with each other through the Hessian potential in Eq.~(\ref{new}). By comparing $\psi$ with $S$, we find that
\begin{eqnarray}
\psi\simeq S.
\end{eqnarray}
This means that the logarithmic formula for the entanglement entropy is a key to producing the warp factor of AdS. Note that we take $g_{\mu\nu}=-\partial_{\mu}\partial_{\nu}S$ with a minus sign in the Ruppeiner metric~\cite{Ruppeiner,Weinhold}. This approach was applied to the classification of classical black holes, although the previous work did not contain the holography concept~\cite{Aman1,Aman2}. Starting with Boltzmann's relation $S=k_{B}\log W$, we obtain $p\propto W= e^{S}=e^{-(-S)}$ . Therefore, the potential is regarded as $-S$, not $S$, and our approach seems to be very different from the past works based on the Ruppeiner metric.

Here, we consider that Eq.~(\ref{ie}) represents the black hole entropy on the classical side as well as the entanglement entropy on the quantum side. We then recall some well-known results associated with the identification of the black hole entropy with the entanglement entropy by a general coordinate transformation. In accordance with the regular coordinate near the event horizon presented in the previous section, let us take
\begin{eqnarray}
z=z_{0}\tanh\left(\frac{l}{z_{0}}\right). \label{tanh}
\end{eqnarray}
We have $z\rightarrow l$ when we take $z_{0}\rightarrow\infty$. This is consistent with the Gaussian case in which $z=\sigma\simeq l$. We find two important aspects for this coordinate transformation. One is about the relationship between $\theta^{2}$ and $l$. By combining Eq.~(\ref{z}) with Eq.~(\ref{tanh}), we obtain
\begin{eqnarray}
\theta^{2}=\frac{1}{z_{0}^{2}\tanh^{2}\left(l/z_{0}\right)}\rightarrow \frac{1}{l^{2}} \label{eq49}
\end{eqnarray}
in the limit of $z_{0}\rightarrow\infty$. This was actually found in a previous work on free fermions by one of the present authors~\cite{Matsueda}. The second aspect is about the entropy formula. By substituting Eq.~(\ref{tanh}) into Eq.~(\ref{ie}), we obtain
\begin{eqnarray}
S=\frac{1}{2}\log\left(z_{0}\sinh\left(\frac{l}{z_{0}}\right)\right)-\frac{1}{2}at^{2}. \label{before}
\end{eqnarray}
This is simply the exact form of the entanglement entropy for CFT${}_{1+1}$ at a finite temperature with an extra time-dependent term if we mutiply this by the central charge. Since there is no physical dimension in the standard information geometry, we must set it appropriately. To recover both the curvature radius $R$ and the overall constant $\kappa$, we require $\kappa=GR$ with the gravity constant $G$ so that the factor $R/G$ is multiplied by $\psi$. Then, the entropy is also multiplied by the same factor $R/G$. With the use of the Brown-Henneaux relation $c=3R/2G$~\cite{Brown}, the logarithmic term in Eq.~(\ref{before}) multiplied by $R/G$ becomes equal to Eq.~(\ref{formula}). As we have already mentioned in the previous section and in Appendix A, the substitution of the regular coordinate into the entropy form is crucial to derive an appropriate scaling formula. It is worth mentioning that quadratic time evolution just after quantum quenching has been observed in previous works, although the coefficient is negative in the present case~\cite{Nezhadhaghighi,AAAL,Liu}.

Now our entropy formula in Eq.~(\ref{ie}) has a negative time-dependent term, by which the information decreases monotonically. Furthermore, this term is dominated by the factor $a=(1/z_{0})^{2}$, which is a function of the event-horizon position. These two important facts suggest that the negative term might play a role in the evaporation of the black hole information~\cite{Bekenstein,Hawking}. Let us consider the situation in which we calculate the black hole entropy by the Bekenstein-Hawking formula $S_{BH}=A/4G$ with horizon area $A$. Here we usually assume the stationary condition that the area of the event horizon is unchanged at the classical level. However, there is Hawking radiation at the quantum level, and the area necessarily decreases. We expect that the negative time-dependent term explains such a dynamical effect. Unfortunatelly, there is a technically difficult problem that the parameter $z_{0}$ should also be dynamical in this case. The time-dependent term is rewritten as $-(1/2)(t/z_{0})^{2}$ and is constant if a linear decrease in $z_{0}$ with $t$ occurs.

It is possible to consider somewhat generic forms of the Hessian potential and the space-time coordinates in order to further examine the time-dependent entropy term. Let us introduce $t$, $x$, and $z$ as
\begin{eqnarray}
t &=& \left(\frac{\theta^{0}}{\theta^{2}-a}\right)^{p}, \\
x &=& \left(\frac{\theta^{1}}{\theta^{2}}\right)^{q}, \\
z &=& \frac{1}{\sqrt{\theta^{2}}},
\end{eqnarray}
where $p$ and $q$ are both nonzero constants. The Hessian potential is defined by
\begin{eqnarray}
\psi&=&\frac{1}{4a}\left\{ (\theta^{2}-a)\log(\theta^{2}-a)-\theta^{2}\log\theta^{2}\right\} \nonumber \\
&& +\frac{1}{2}\frac{q}{2q-1}\left(\frac{\theta^{1}}{\theta^{2}}\right)^{2q}\theta^{2} \nonumber \\
&& -\frac{1}{2}\frac{p}{2p-1}\left(\frac{\theta^{0}}{\theta^{2}-a}\right)^{2p}(\theta^{2}-a).
\end{eqnarray}
Then, we again obtain the BTZ metric, and the entropy formula is represented as
\begin{eqnarray}
S=\frac{1}{4}\log\left(z_{0}\sinh\left(\frac{l}{z_{0}}\right)\right)-\frac{p}{2}at^{2}.
\end{eqnarray}
We find that the leading logarithmic term is invariant for any $p$ and also find that the sign of the time-dependent term is dominated by the sign of $p$. The presence of the coordinate-dependent entropy is reminiscent of the Unruh effect, and thus may indicate some radiation~\cite{Fulling,Davies,Unruh}. The entropy formula does not depend on the parameter $q$. For $q$, we simply assume $q=p$ to ensure the symmetry between $\theta^{1}$ and $\theta^{0}$. The time coordinate is then given by
\begin{eqnarray}
t=\left(\frac{\theta^{0}}{\theta^{2}-a}\right)^{p}=\left(\frac{z^{2}}{1-az^{2}}\right)^{p}\left(\theta^{0}\right)^{p}.
\end{eqnarray}
The factor $(z^{2}/(1-az^{2}))^{p}$ diverges at $z=z_{0}$ if $p$ is positive. Then, this choice correctly represents the time evolution near the event horizon. Therefore, the time-dependent term should be negative, and this analysis also seems to support the radiation scenario.

\section{Legendre Transformation and Duality}

As shown in the previous sections, there are many other sets of Hessian potentials and canonical parameters that produce the same BTZ metric. This ambiguity seems to be consistent with the renormalization group concept in the sense that the classical side represents the universal feature of various quantum models. Among them, we find two special representations that are dual in terms of the Legendre transformation. According to Eqs.~(\ref{eta}) and (\ref{varphi}), the duality transformation for Eq.~(\ref{new}) is given by
\begin{eqnarray}
\eta_{0} &=& \frac{\theta^{0}}{\theta^{2}-a} , \\
\eta_{1} &=& -\frac{\theta^{1}}{\theta^{2}} , \\
\eta_{2} &=& -\frac{1}{4a}\left\{ \log(\theta^{2}-a)-\log\theta^{2} \right\} \nonumber \\
&& -\frac{1}{2}\frac{(\theta^{0})^{2}}{(\theta^{2}-a)^{2}}+\frac{1}{2}\frac{(\theta^{1})^{2}}{(\theta^{2})^{2}}.
\end{eqnarray}
They are inverted to the original canonical parameters as follows:
\begin{eqnarray}
\theta^{0} &=& a\frac{e^{V}}{1-e^{V}}\eta_{0} , \\
\theta^{1} &=& -a\frac{1}{1-e^{V}}\eta_{1} , \\
\theta^{2} &=& a\frac{1}{1-e^{V}}, \label{th2}
\end{eqnarray}
where
\begin{eqnarray}
V(\eta)=-4a\left(\eta_{2}+\frac{1}{2}(\eta_{0})^{2}-\frac{1}{2}(\eta_{1})^{2}\right). \label{p3}
\end{eqnarray}
The dual potential $\varphi$ is given by
\begin{eqnarray}
\varphi &=& -\theta^{\alpha}\eta_{\alpha}-\psi \nonumber \\
&=& \frac{1}{4}\log(\theta^{2}-a)+\frac{1}{2}a\frac{(\theta^{0})^{2}}{(\theta^{2}-a)^{2}} \nonumber \\
&=& -\frac{1}{4}\log(1-e^{V})+\frac{1}{4}\log a + \frac{1}{4}V + \frac{1}{2}a(\eta_{0})^{2}. \label{vphi}
\end{eqnarray}
Using the dual parameters, the space-time coordinates and the scale parameter are represented as
\begin{eqnarray}
t &=& \eta_{0}, \label{ttt} \\
x &=& -\eta_{1}, \label{xxx} \\
z &=& \sqrt{\frac{1}{a}(1-e^{V})} . \label{eV0}
\end{eqnarray}
Equation (\ref{eV0}) is rewritten as
\begin{eqnarray}
e^{V}=f=1-az^{2}. \label{eV}
\end{eqnarray}
By using the parameters $t$, $x$, and $z$, the dual potential is given by
\begin{eqnarray}
\varphi= -\frac{1}{4}\log\left(\frac{z^{2}}{1-az^{2}}\right)+\frac{1}{2}at^{2}.
\end{eqnarray}
Our result is a direct extension of the well-known construction of the pure AdS metric to finite-$z_{0}$ cases~\cite{Shima}. The potential form in Eq.~(\ref{vphi}) is essentially equal to the free energy for free bosons, which is a typical CFT model. Thus, $\varphi$ can be regarded as the entanglement thermodynamic potential for the entanglement Hamiltonian. In this case, $V$ (or otherwise $\eta_{2}$) represents the energy scale of the entanglement Hamiltonian and $a$ is related to the inverse temperature of the black hole radiation such as the Rindler or Hawking temperature.

Therefore, the two different representations are holographically dual objects. The former representation is based on the BTZ black hole on the bulk side and the latter is based on the CFT at the boundary of our space-time in terms of the AdS/CFT correspondence. Let us again examine the leading logarithmic term of the entropy formula:
\begin{eqnarray}
S=-\frac{1}{2}\log\left\{\frac{1}{\frac{\beta}{\pi}\sinh\left(\frac{2\pi l}{\beta}\right)}\right\}-\frac{1}{2}at^{2}.
\end{eqnarray}
The expression inside of the logarithm is equal to the two-point correlator of scalar field theory at a finite temperature. This theory was actually used by Calabrese and Cardy from the viewpoint of the CFT~\cite{Calabrese1,Calabrese2,Calabrese3}.

For $a\rightarrow 0$ and $\eta_{0}=\eta_{1}$ in Eq.~(\ref{th2}), the duality associated with the energy scale is simply represented as
\begin{eqnarray}
\theta^{2}\eta_{2}=\frac{1}{4}.
\end{eqnarray}
This equality is reminiscent of the self-dual property such as the Kramers-Wannier symmetry in classical spin models.

Finally, we briefly mention that the dual representation also contains the physics in the Rindler space. We have $V\le 0$ for $z\ge 0$ in Eq.~(\ref{eV0}). At the boundary of our space ($z=0$), we find from Eq.~(\ref{p3}) that
\begin{eqnarray}
2\eta_{2}+t^{2}-x^{2}=0.
\end{eqnarray}
This equality means the hyperbolic worldline, and by varying the magnitude of $\eta_{2}$ ($\ge 0$) the worldlines cover the whole space of the left and right Rindler wedges. The emergence of the Rindler physics from the quantum entanglement may have some connection to the physics of the topological black hole~\cite{Emparan}.

\section{Summary}

We examined the BTZ black hole in terms of the information geometry and AdS/CFT correspondence. Our conclusion is that the information geometry can capture very important holographic properties in the AdS/CFT correspondence. We found that the BTZ black hole is actually dual to the finite-temperature CFT${}_{1+1}$. To prove this, we obtained the exact Hessian potential and also found that it is crucial to select coordinates that are regular at the event horizon. Furthermore, we discussed the role of the Legendre transformation in the representation of the Hessian potential. Even though the potential form changes after the transformation, the resultant metric is invariant. Therefore, the renormalization-group nature of the holography originates from this transformation in terms of the information geometry. We found that for a particular choice the potential is closely related to the free-energy form for the entanglement Hamiltonian on the quantum side. In this representation, we found that the two-point correlator of the free scalar field behaves as the radial coordinate, consistent with the Calabrese-Cardy formula. Therefore, the Legendre transformation seems to also be related to UV/IR correspondence. The present result is expected to shed new light on recent works on gravitational dynamics emerging from entanglement entropy~\cite{Blanco,Casini,Wong,Takayanagi2,Takayanagi3,Faulkner,Banerjee,Nima}. Although the present study did not contain any perturbation from the stationary state, it is straightforwardly possible to extend the present theory to such dynamical cases. This will be one of our future works. To obtain a more complete picture, it is necessary to find the explicit form of $\theta^{\alpha}$ with the use of various quantum models. We have already found that $\theta^{2}=1/l^{2}$ in the free-fermion case, but this is only one example. We must find other examples to extend our discussion.

HM acknowledges Benjamin Meiring, Jonathan Shock, Yoichiro Hashizume, and Isao Maruyama for their fruitful comments and discussion. This work was supported by JSPS KAKENHI Grant Numbers 15K05222 and 15H03652.

\appendix

\section{Transformation of the Thermal to Entanglement Entropy: Roles of General Coordinate Transformation in the Entropy Formula}

We present a well-known example of the close relationship between black hole thermodynamics and entanglement, and show how the general coordinate transformation plays a role in finding their relationship. Another example is the topological black hole, which also contains the same physics~\cite{Emparan}.

In CFT${}_{1+1}$, the thermal entropy of a system defined in the spatial region $[\xi_{1},\xi_{2}]$ is
\begin{eqnarray}
S=\frac{\pi}{3}cT\left(\xi_{2}-\xi_{1}\right). \label{a0}
\end{eqnarray}
Here we introduce the coordinate transformation
\begin{eqnarray}
\xi=\frac{1}{\kappa}\log\left(\kappa X\right), \label{xx}
\end{eqnarray}
where $X>0$ for $-\infty<\xi<\infty$. This general coordinate transformation truncates the negative $X$ region and thus is related to entanglement. We regard $T$ as a characteristic temperature such as the Rindler temperature, which is defined by
\begin{eqnarray}
T=\frac{\kappa}{2\pi}.
\end{eqnarray}
Then, we obtain the logarithmic entropy formula for a half-filled 1D chain as
\begin{eqnarray}
S=\frac{\pi}{3}c\frac{\kappa}{2\pi}\frac{1}{\kappa}\log\left(\frac{X_{2}}{X_{1}}\right)=\frac{c}{6}\log\left(\frac{L}{\epsilon}\right),
\end{eqnarray}
where we have taken $X_{1}=\epsilon$ and $X_{2}=L$.

The general coordinate transformation in Eq.~(\ref{xx}) appears in the theory of Rindler black holes. Here, $X$ is the spatial coordinate of Minkowski spacetime, and $\xi$ is the coordinate of an observer in the Rindler wedge. The Rindler coordinate is defined as a limiting case of the Schwarzschild black hole. Therefore, Eq.~(\ref{a0}) behaves as the black hole entropy in the sense that the confinement prohibits any access to information outside the wedge.

\section{Codazzi Equations}

In this appendix, we show how to derive Eq.~(\ref{time}) from the Codazzi equations.

The condition that $g_{\mu\nu}$ is a Hessian form is equivalent to the condition that the metric satisfies the following Codazzi equations:
\begin{eqnarray}
\partial_{\lambda}g_{\mu\nu}=\partial_{\mu}g_{\lambda\nu},
\end{eqnarray}
where we find that
\begin{eqnarray}
\partial_{\lambda}g_{\mu\nu}=\partial_{\lambda}\partial_{\mu}\partial_{\nu}\psi=2\Gamma_{\lambda\mu\nu},
\end{eqnarray}
Using the Christoffel symbol $\Gamma_{\lambda\mu\nu}$, and the above equations simply correspond to the symmetry for the exchange of the order of the indices. For the BTZ metric, let us assume
\begin{eqnarray}
t &=& t(\theta^{0},\theta^{2}) , \\
x &=& \frac{\theta^{1}}{\theta^{2}} , \\
z &=& \frac{1}{\sqrt{\theta^{2}}}.
\end{eqnarray}
Here we assume that we have already obtained appropriate forms of $x$ and $z$ by elementary study of the Gaussian distribution. Then, we have $f(z)/z^{2}=\theta^{2}-a$, and taking
\begin{eqnarray}
t_{\mu}=\frac{\partial t}{\partial\theta^{\mu}},
\end{eqnarray}
we obtain
\begin{eqnarray}
ds^{2} &=& \frac{1}{z^{2}}\left(-f(z)dt^{2}+\frac{dz^{2}}{f(z)}+dx^{2}\right) \nonumber \\
&=& -(\theta^{2}-a)(t_{0}d\theta^{0}+t_{2}d\theta^{2})^{2} \nonumber \\
&& + \frac{1}{\theta^{2}}(d\theta^{1})^{2}-2\frac{\theta^{1}}{(\theta^{2})^{2}}d\theta^{1}d\theta^{2} \nonumber \\
&& + \frac{1}{(\theta^{2})^{2}}\left(\frac{(\theta^{1})^{2}}{\theta^{2}}+\frac{\theta^{2}}{4(\theta^{2}-a)}\right)(d\theta^{2})^{2} .
\end{eqnarray}
Therefore, we obtain
\begin{eqnarray}
g_{00}&=&-(\theta^{2}-a)(t_{0})^{2} , \\
g_{01}&=&0 , \\
g_{02}&=&-(\theta^{2}-a)t_{0}t_{2} , \\
g_{11}&=&\frac{1}{\theta^{2}} , \\
g_{12}&=&-\frac{\theta^{1}}{(\theta^{2})^{2}} , \\
g_{22}&=&\frac{1}{(\theta^{2})^{2}}\left(\frac{(\theta^{1})^{2}}{\theta^{2}}+\frac{\theta^{2}}{4(\theta^{2}-a)}\right) \nonumber \\
&& -(\theta^{2}-a)(t_{2})^{2} ,
\end{eqnarray}
where we have the symmetry $g_{\mu\nu}=g_{\nu\mu}$. Now we have a total of nine Codazzi equations, but only two of them are nontrivial equations. By introducing the new notation
\begin{eqnarray}
t_{\mu\nu}=\frac{\partial^{2}t}{\partial\theta^{\mu}\partial\theta^{\nu}},
\end{eqnarray}
we find that these two are
\begin{eqnarray}
\partial_{0}g_{20}=\partial_{2}g_{00} , \\
\partial_{0}g_{22}=\partial_{2}g_{02} ,
\end{eqnarray}
which are respectively given by
\begin{eqnarray}
(\theta^{2}-a)(t_{2}t_{00}-t_{0}t_{02})-(t_{0})^{2}&=&0 , \\
(\theta^{2}-a)(t_{0}t_{22}-t_{2}t_{02})+t_{0}t_{2}&=&0 .
\end{eqnarray}
Furthermore, when the variables are separated as
\begin{eqnarray}
t=h(\theta^{0})k(\theta^{2}),
\end{eqnarray}
they become
\begin{eqnarray}
(\theta^{2}-a)\frac{k^{\prime}}{k} &=& -p , \\
\frac{h^{\prime\prime}h-(h^{\prime})^{2}}{(h^{\prime})^{2}} &=& -\frac{1}{p} ,
\end{eqnarray}
where $p$ is a constant. We obtain
\begin{eqnarray}
k(\theta^{2}) &=& \frac{1}{(\theta^{2}-a)^{p}} , \\
h(\theta^{0}) &=& (\theta^{0})^{p} .
\end{eqnarray}
Finally, the time coordinate is represented as
\begin{eqnarray}
t=\left(\frac{\theta^{0}}{\theta^{2}-a}\right)^{p}.
\end{eqnarray}
This result indicates that there are many possibilities for $\psi$ and the coordinates to finally obtain the same BTZ metric. Here, we have started with a modification of the Gaussian distribution, but more general transformations are expected to exist.

\end{document}